# High Kondo temperature ($T_K \sim 80$ K) in self-assembled InAs quantum dots laterally coupled to nanogap electrodes

K. Shibata[a)] and K. Hirakawa[b)]

*Institute of Industrial Science and Institute for Nano Quantum Information Electronics, University of Tokyo, 4-6-1 Komaba, Meguro-ku, Tokyo 153-8505, Japan;*
*CREST-JST, 4-1-8 Honcho, Kawaguchi, Saitama 332-0012, Japan*

We have fabricated single electron tunneling structures by forming nanogap metallic electrodes directly upon single self-assembled InAs quantum dots (QDs). The fabricated samples exhibited clear Coulomb blockade effects. Furthermore, a clear Kondo effect was observed when strong coupling between the electrodes and the QDs was realized using a large QD with a diameter of ~100 nm. From the temperature dependence of the linear conductance at the Kondo valley, the Kondo temperature $T_K$ was determined to be ~ 81 K. This is the highest $T_K$ ever reported for artificial quantum nanostructures.

The spin-half Kondo effect in single electron transistors (SETs) is a coherent many-body state comprising a localized unpaired spin on the quantum dot (QD) which interacts with the opposite spins of conduction electrons in the electrodes [1]. The Kondo effect appears as an enhanced zero-bias conductance (Kondo resonance) when the temperature, $T$, is below a characteristic temperature called the Kondo temperature, $T_K$. Since SETs have large flexibility for tuning various system parameters, the Kondo effect in SETs has been attracting much attention in studies of many-body spin interactions in a controlled environment [2].

The Kondo effect in SETs has been observed in lithographically defined semiconductor QDs [2,3], carbon nanotubes [4], and single molecule transistors [5]. However, except for the case of single molecule transistors ($T_K \sim$10-25 K) [5], the Kondo effect in SETs is usually observed only at very low temperatures (typically <1 K). It is known that $k_B T_K$ ($k_B$: the Boltzmann constant) is limited by the quantum mechanical coupling energy between the QD and the electrodes, $h\Gamma$. On the other hand, electrons are no longer localized in the QD if $h\Gamma$ exceeds the charging energy, $E_c$, or the orbital quantization energy, $\Delta E$. Therefore, $T_K$ is governed by the relationship, $k_B T_K \leq h\Gamma < E_c, \Delta E$ [2].

We have studied electron tunneling through single self-assembled InAs QDs coupled to metallic nanogap electrodes [6-11]. Recently, we observed a spin-half Kondo effect with $T_K$ as high as ~10 K [8-11]. However, considering that InAs QDs have large $E_c$ and $\Delta E$ of the order of 10 meV [6-7], they may exhibit much higher $T_K$ if stronger coupling with the electrodes is realized.

In this work, we have concentrated our efforts on the realization of high Kondo temperatures in the InAs QD system. We have fabricated single QD SET structures with low tunnel junction resistances by carefully depositing metallic electrodes with narrow gaps directly upon single self-assembled InAs QDs with diameters of ~100 nm. The fabricated samples exhibited clear Coulomb blockade effects. Furthermore, the Kondo effect was clearly observed in Coulomb diamonds for odd electron occupancy in the QD. The Kondo temperature determined from transport data exceeded 80 K, which is significantly higher than the previously reported values in SETs [1-5, 8-11]. This high Kondo temperature is due to strong QD-electrode coupling and large charging/orbital-quantization energies in the present self-assembled InAs QD structures.

Self-assembled InAs QDs were grown by molecular beam epitaxy on (100)-oriented $n^+$-GaAs substrates. After successively growing a 100 nm-thick $Al_{0.3}Ga_{0.7}As$ barrier layer and a 200 nm-thick undoped GaAs buffer layer, InAs QDs were grown at 500 °C. The InAs coverage of ~ 4-6 ML was used to obtain a mixed phase of large and small QDs, as shown in Fig. 1(a). The n-type substrate was used as a backgate electrode. The pattern of the nanogap electrodes was defined by electron beam lithography in PMMA resist. Metallic electrodes were formed directly upon the uncapped QDs by successively depositing 10 nm-thick Ti and 10 nm-thick Au layers. Prior to the metal deposition, the sample wafers were dipped into buffered hydrofluoric acid for 6 s to remove native oxides on the QD surfaces. The proper design of the gap and the width of the electrodes makes it possible to realize a situation in which most of the conductive junctions contain single QDs. Approximately 10 percent of the fabricated junctions showed reasonable conductances at 300 K. Details of the nanofabrication process have been reported elsewhere [6,7].

From our previous work, it is known that the junction resistances of the QD SETs strongly depend on the size of the InAs QDs; i.e., the low-temperature junction resistances for large QDs with diameters > 90 nm are three orders of magnitude lower than those for smaller QDs with diameters < 80 nm [6]. This fact means that strong coupling between the QD and the electrodes can be achieved when a large InAs QD is used as a Coulomb island. Therefore, we chose a rather large QD with a diameter of ~100 nm in this work. Although the origin of the strong QD-size dependence of the tunneling resistance is not clear at present, we speculate that intermixing between In and Ga atoms during crystal growth plays an important role. It is known that, when QDs are grown around 500 °C, a large composition gradient from GaAs to InAs forms up to 3-5 nm above the wetting layer because of the intermixing effect [12]. Due to this composition gradient, the electron wave functions are pushed toward the apex of the QDs [13]. For the case of small QDs

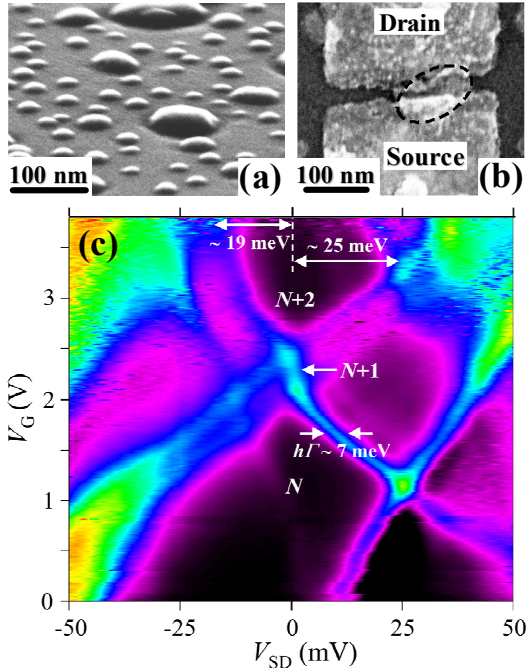

FIG 1 (a) An SEM image of the sample used in the experiment. The sample contains a mixed phase of large and small InAs QDs. (b) An SEM image of the fabricated sample. The dashed circle in the figure denotes the InAs quantum dot in the gap. (c) Coulomb stability diagram obtained by plotting the differential conductance $dI/dV_{SD}$ as a function of $V_{SD}$ and $V_G$.

with typical heights of ~10 nm, a large portion of the QD is alloyed with Ga and electrons are localized near the apex. Therefore, the coupling between the electrons in the QD and the electrodes is weak, as shown in the inset of Fig. 3. For the case of large QDs with diameters of ≥ 100 nm, on the other hand, a larger portion of the QD is made of pure InAs and more electrons accumulate. As the number of electrons in the QD increases, electrons occupy higher energy shells and their wavefunctions become more extended in space [7], as illustrated schematically in the inset of Fig. 3. Since the tunneling resistance in the present QD SETs is determined by the coupling between the metallic electrodes and the electron wave function in the QD [7,8], the resistance is smaller for large QDs.

Figure 1(b) shows a scanning electron microscope (SEM) image of a sample which exhibited a very low junction resistance (~30 kΩ) at 300 K. As seen in the figure, metallic electrodes with a ~20 nm-wide nanogap separation are bridged by a single elliptic InAs QD with a diameter of 60/120 nm (the dashed oval in Fig. 1(b)). Since the contact area for the source electrode is larger than that for the drain electrode, the tunneling resistances are asymmetric.

Figure 1(c) shows the Coulomb stability diagram obtained by measuring the differential conductance, $dI/dV_{SD}$, as a function of the source-drain voltage, $V_{SD}$, and the gate voltage, $V_G$, at 4.2 K. As seen in the figure, the boundaries of the Coulomb diamonds are broadened and not straight, reflecting strong coupling between the QD and the electrodes. Although the number of electrons in the QD, $n$, could not be determined, we can tell the parity from the size of the Coulomb diamonds. The diamonds are larger for the regions labeled $N$ and $N+2$. Larger diamonds indicate that these regions contain even numbers of electrons ($N$ is an even number) and have larger addition energies that consist of both the charging energy and the orbital quantization energy in the QD. On the other hand, the diamond labeled $N+1$ is smaller, because the addition energy is only due to the charging energy.

The inset of Fig. 2(a) shows a blow-up of the Coulomb diamond for $n = N+1$ (odd). The asymmetric shape of the Coulomb diamond is due to the asymmetric coupling between the QD and the source/drain electrodes [14]. As seen in the inset of Fig. 2(a), the Coulomb diamond for $n = N+1$ shows a clear resonant enhancement in zero-bias conductance. This behavior arises from the formation of a spin-singlet state between an unpaired electron in the QD and an electron with the opposite spin in the electrodes; i.e., the spin-half Kondo effect. From the inset of Fig. 2(a), the charging energy, $E_c$, in this sample is estimated to be ~ 7-8 meV. From the Coulomb diamond for $n = N+2$, the addition energy, $E_{add}$, for $n = N+2$ is determined to be ~ 22 meV. The orbital quantization energy, $\Delta E$, is, therefore, estimated to be $E_{add} - E_c$ ~ 14-15 meV. To determine the coupling energy between the QD and the electrodes, $h\Gamma$, we set $V_G$ in such a way that the energy levels corresponding to the paired peaks were empty. Then, the width of the peak in the plot of $dI/dV_{SD}$ versus $V_{SD}$ gives the bare level width $h\Gamma \sim 7$ meV (see Fig. 1(c)) [2]. The coupling energy to the source (drain) electrodes can also be obtained from $h\Gamma_{S(D)} \sim h\Delta I_{S(D)}/e$, where $\Delta I_{S(D)}$ is the increase in current when an additional transport channel is opened in resonance with the source (drain) contact [14]. This yields $h\Gamma_S \sim 4.3$ meV and $h\Gamma_D \sim 2.5$ meV (not shown in Fig. 1(c)), which is consistent with the total coupling energy $h\Gamma = h\Gamma_S + h\Gamma_D \sim 7$ meV. The large obtained $h\Gamma$ reflects strong coupling between the QD and the metallic electrodes.

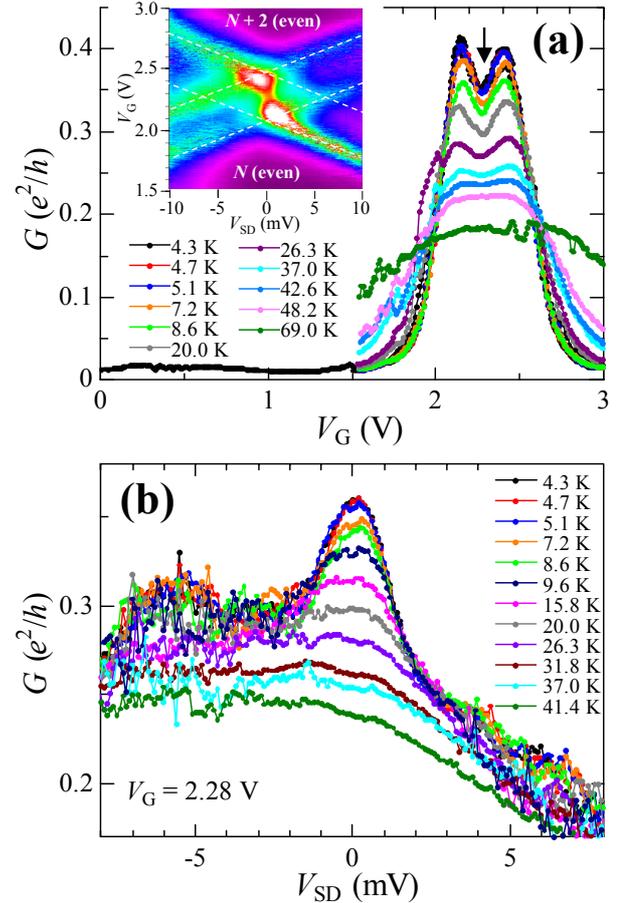

FIG 2: Temperature dependence of the conductance at the middle of the Kondo valley ($n = N+1$). (a) Linear conductance spectra taken at various temperatures by applying $V_{SD} = 0.1$ mV and sweeping the backgate voltage, $V_G$. The inset shows a blow-up of the color-coded differential conductance around the Kondo valley. (b) Differential conductance as a function of $V_{SD}$ measured at the middle of the Kondo valley ($V_G = 2.28$ V) at various temperatures.

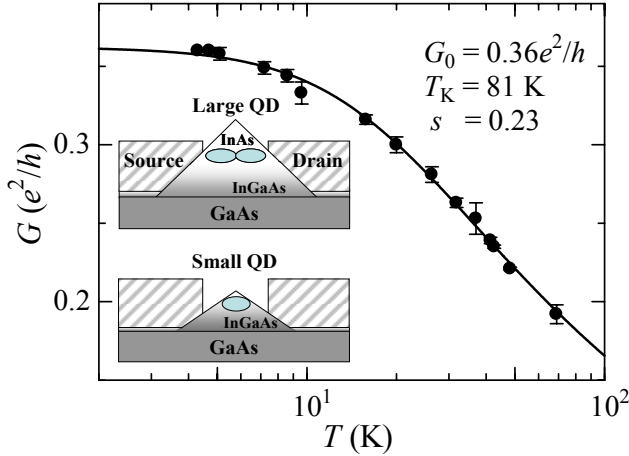

FIG. 3 Temperature dependence of the conductance at the middle of the Kondo valley ($n = N+1$). The solid line denotes the fit of the empirical function to the data. The inset shows schematic illustrations of the sample geometries and the electron wave functions for large and small QD samples.

Figure 2(a) shows the linear conductance spectra measured at various temperatures from 4.2 K to 69 K. The conductance, $G$, at the middle of the $n = N+1$ Coulomb valley (the Kondo valley; indicated by an arrow) grows with decreasing temperature and saturates at 0.36 $e^2/h$ at low temperatures. The observed saturated conductance is smaller than that expected for the unitary limit ($2e^2/h$), which can be explained by the asymmetric coupling between the QD and the source/drain electrodes in this sample [9]. In Fig. 2(b), $dI/dV_{SD}$ is plotted as a function of $V_{SD}$ at $V_G = 2.28$ V for various temperatures. The Kondo peak around $V_{SD} = 0$ V grows with decreasing temperature. Although the background conductance due to the falling edges of the adjacent tunneling peaks hinders precise determination of the width of the Kondo peak, $w$, it is roughly estimated to be $ew > 4$ meV. The relationship $k_B T_K \cong ew$ implies that the Kondo temperature $T_K$ is higher than 45 K.

Figure 3 shows the temperature dependence of the linear conductance at the middle of the Kondo valley ($V_G = 2.28$ V). The solid line is a fit of the empirical function [9], $G(T) = G_0[1/(1+(2^{1/s} - 1)(T/T_K)^2)]^s$, to the experimental data. A good fit was obtained when the parameters were set as $G_0 \sim 0.36$ $e^2/h$, $s \sim 0.23$, and $T_K \sim 81$ K. This Kondo temperature is the highest value ever reported for artificial quantum nanostructures. The fact that $k_B T_K$, $h\Gamma$, $\Delta E$, and $E_c$ have similar values implies that the observed Kondo temperature is close to the upper limit which can be realized in our InAs QD SET structures.

In summary, we have fabricated single electron tunneling structures by forming nanogap metallic electrodes directly upon single self-assembled InAs QDs. The fabricated samples exhibited clear Coulomb blockade effects. Furthermore, a clear Kondo effect was observed when strong coupling between the electrodes and the QDs was realized by using a large QD with a diameter of ~100 nm. From the temperature dependence of the linear conductance at the Kondo valley, the Kondo temperature $T_K$ was determined to be ~ 81 K. This is the highest $T_K$ ever reported for artificial quantum nanostructures. This high Kondo temperature is due to strong QD-electrode coupling and large charging/orbital-quantization energies in the present self-assembled InAs QD structures.

We thank H. Sakaki, Y. Arakawa, S. Tarucha, T. Machida, A. Oiwa, K. Hamaya, and M. Jung for fruitful discussions, and S. Ishida for technical support. This work was partly supported by the Japan Science and Technology Corporation (CREST), Grants-in-Aid from the Japan Society for Promotion of Science (No. 18201027 and No. 19560338), and the Specially Coordinated Fund from MEXT (NanoQuine).

___________________________
a) Electronic address: kshibata@iis.u-tokyo.ac.jp
b) Electronic address: hirakawa@iis.u-tokyo.ac.jp